\documentclass[twocolumn,showpacs,preprintnumbers,amsmath,amssymb]{revtex4-1}
\usepackage{epsfig,amsopn}
\usepackage{graphicx}
\usepackage{epstopdf}
\usepackage{sidecap}
\usepackage{amsmath,amssymb}
\usepackage{amsthm}
\usepackage{enumerate}
\usepackage{bbold}
\newcommand\bea{\begin{eqnarray}}
\newcommand\eea{\end{eqnarray}}
\newcommand\beq{\begin{equation}}  
\newcommand\eeq{\end{equation}}

\newcommand{\bib}{\bibitem}
\newcommand{\cosec}{\operatorname{cosec}}

\begin{document}
\title{\bf{Transport and STM  studies  of hyperbolic surface states  of topological insulators}}
\author{ Udit Khanna, Saurabh Pradhan and Sumathi Rao}
\affiliation{Harish-Chandra Research Institute, Chhatnag Road, 
Jhusi, Allahabad 211 019, India.}
\begin{abstract}
Motivated by the transmission of topological surface states through atomic scale steps,
we study the transport of gapless Dirac fermions on hyperbolic  surfaces.  
We confirm that,  independent of the curvature of the hyperbolae and the sharpness of
the corners, no backward scattering takes place and transmission 
of the topological surface states is completely independent of the geometrical shape
(within the hyperbolic model) of the surface. The density of states of the electrons, however, 
shows a dip at concave step edges which    
can be measured by an STM tip. We also show that the tunneling conductance
measured by a   polarized scanning tunneling probe exhibits an unconventional dependence
on the polar and azimuthal angles of the magnetization of the tip 
as a function of the  curvature of the surface and the sharpness of the edge.
\end{abstract} 
\pacs{73.20.-r, 73.63.Nm, 71.10.Pm}
\maketitle
\section{Introduction} 

Interest in topological insulators (TIs) continues to remain high
ever since their prediction\cite{theory} and discovery\cite{konig}.  The hallmark
of these materials is their topologically non-trivial nature, due to
which they have gapless linearly dispersing edge states, although they
are insulating in the bulk\cite{reviews}.  Such edge states are familiar in the two-dimensional 
quantum Hall system\cite{qhereviews},  but in TIs, they arise even in 
higher dimensional systems and even in the absence of
time-reversal symmetry breaking. 

For two-dimensional TIs, the
edge states consist of two counter-propagating  modes, with opposite
spin projections. The correlation of the spin of the electron with its direction
of motion is the central feature of these states.
The existence of these states have been verified experimentally
by measurements  in the HgTe quantum well structures\cite{konig}. 
 There have also been several theoretical studies \cite{vari, das,soori}, studying
 consequences of the spin projection and the helical  nature  of the edge states,
 although there has been no direct measurement of the spin of the edge states.

The three dimensional TIs  come in two classes, strong and
weak, with even and odd number of Dirac cones on its surface. For strong
TIs, the existence of an odd number of Dirac cones is topologically protected and
it has been found that the surface states of strong TIs  such as $HgTe, Bi_2Se_3$ 
and $Bi_2Te_3$  can be described by a single Dirac electron.
These surface states are robust against perturbations that do not break time-reversal
symmetry. The electrons also have the feature that their spin is `locked'  to the
momentum of the electron, which is what leads to a complete absence of back-scattering,
because the spin of the electrons at momentum ${\bf k}$ is orthogonal to the spin
of the electrons at momentum $-{\bf k}$. The combination of topological stability
and absence of backward scattering leads to the prediction that the surface states 
of TIs wrap the surface of the TI and are impervious to the existence of 
surface defects. This has been confirmed by experiments which not only have
confirmed the absence of backward scattering\cite{tzhang, roushan}, 
but more recently have also
shown full transmission through atomic step edges\cite{seo}.   This would imply that
transport through these surface states is independent of the geometrical shape
of the surface.

With a few  notable exceptions\cite{imura1,imura2,imura3}, most of the theoretical work on TIs has been 
restricted to planar surfaces and straight line edges. For instance,
the original derivation of the edge states was for samples with a single straight 
edge. This was extended to the case for finite strips where it
was shown that there was some interference between the two edges\cite{zhou}.
However, there had been no direct study of what happens when two sharp edges meet
one another. Since spin projection is tied to the direction of motion, it is also 
not clear how the spin current changes if there are sharp edges. Similarly,
although there have  been some studies\cite{jiang,takahashi,diptiman,krishnendu} 
of what happens at junctions of two 
3DTIs, the extension of the gapless Dirac state over curved surfaces has yet to be
demonstrated.

A step in this direction was recently taken by Takane and Imura\cite{takane}, who
introduced a hyperbolic system to treat the $90^{\rm o}$ step edge, and 
by introducing  appropriate curvilinear coordinates,  they could show that 
no reflection takes place at the $90^{\rm o}$ step edge, and transmission was perfect,
although there was a  
sharp change in the expectation value of spin in the close vicinity of the step. 
In this paper, we generalise their work to step edges of arbitrary angles,
and show that independent of the curvature of the hyperbolae and the sharpness of
the corners, no backward scattering takes place and the transmission 
of the topological surface states is completely independent of the geometrical shape
(within the hyperbolic model) of the surface.  Moreover, we 
study  how the density of states (DOS)
and the spin DOS behave as a function of the curvature and the
sharpness of the edge of any sample.  We find that the DOS
shows a dip at the concave edges of the sample.
We also  compute the tunneling conductance measured by a polarized scanning tunneling microscope,
as a function of the curvature of the surface and the sharpness of the edges and show
that the STM conductance has a non-trivial dependence on the curvature angle $\phi$ and
an unconventional dependence on the polar and azimuthal angles of the tip, which
are not displayed by a planar TI surface.

\section{The model and analysis}
We start with the continuum model of a strong anisotropic TI in 3D,
given by the Hamiltonian\cite{hzhang}
\beq
H_{\rm bulk} = \begin{pmatrix}{} m({\bf p})\tau_z +A_z  p_z \tau_x & (A_x p_x -  i A_y p_y)\tau_x \\
                                  (A_x p_x + iA_y p_y) \tau_x  &   m({\bf p})\tau_z - A_z p_z \tau_x\end{pmatrix}~.
\eeq
The parameters $A_x,A_y,A_z$ and $m({\bf p})$
can be determined for specific materials by comparing with the {\it ab initio} calculations of 
the effective model of 3DTI\cite{hzhang} . Here, the $2 \times 2$ matrix represents the spin degrees
of freedom and the orbital degrees of freedom are represented by the Pauli matrices $\tau_i$.
The mass term $m({\bf p}) = m_0 +m_2(p_x^2+p_y^2+p_z^2)$ in this  3D Dirac Hamiltonian 
is momentum dependent. For the topological insulator, we need to take $m_0>0$ and $m_2<0$.
Assuming the system to be translationally invariant in the $x$ direction, so that $p_x$ is a good
quantum number, it is straightforward to derive the surface states on either the $xy$ plane or
the $xz$ plane. However, it is not obvious what happens close to the corners. Takane and
Imura\cite{takane} studied the question of whether or not there exists reflection at the
corners for non-zero $p_x$ by assuming that the surface could be represented by a rectangular
hyperbolic model using curvilinear coordinates. In this paper, we assume that the surface of the
TI can be an arbitrary hyperbolic surface as shown in Fig. (1a).  Note that we are representing
a concave surface - the TI is in the shaded region as shown in the figure.
The curve in Fig. (1b)
can be described by the equation 
\beq
(az)^2 - y^2 = R^2
\eeq
where the angle between the two asymptotic surfaces is governed by the (curvature) parameter $a$
and the sharpness at the edge is governed by the (sharpness) parameter $R$. The 3D TI is 
translationally invariant in the $x$-direction.  A separable coordinate system can now be defined
as follows - 
\bea
y&=& -r\cos\phi +y_0(\phi), \nonumber \\
z &=& -r \sin\phi +z_0(\phi),
\eea
where $y_0$ and $z_0$ define the curve and are given by 
\bea 
y_0 &=& \frac{-aR \cot \phi}{\sqrt{1-a^2\cot^2\phi}} \nonumber \\
{\rm and} ~~z_0 &=& \frac{R}{a} \frac{1}{\sqrt{1-a^2\cot^2\phi}}~.
\eea

\begin{figure}
\epsfig{figure=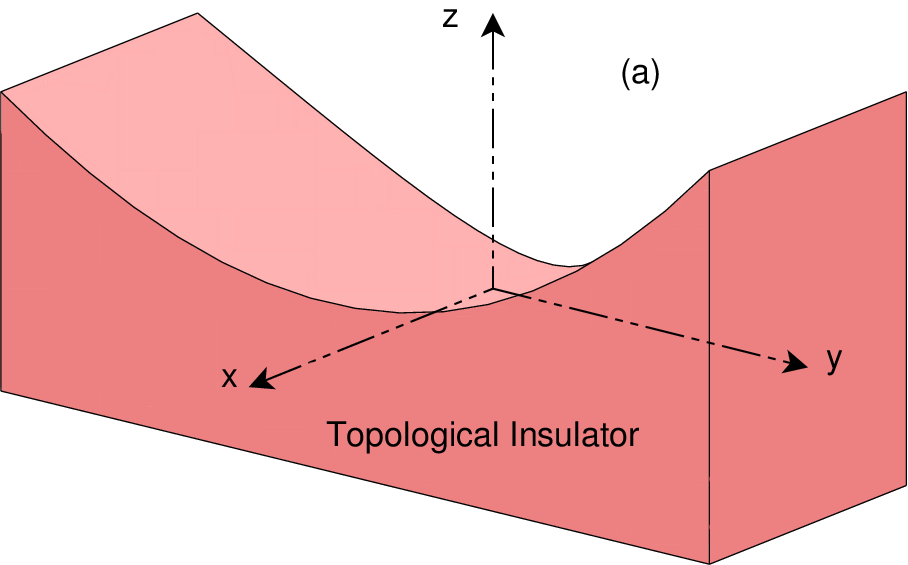,width=5cm,height=3.5cm}
\epsfig{figure=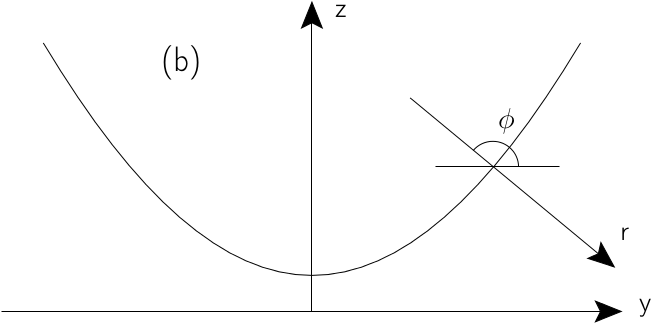,width=7cm,height=3.5cm}
\caption{The surface is shown (above)  in Fig.(1a) with the TI filling the concave part and the
curvilinear coordinate system is shown (below) in Fig.(1b) as described in the text.}\label{coordinate}
	\end{figure}
We draw  normals to the surface  as shown in Fig. (1b) and define $\phi$ as the angle
between the normal and the $y$-axis, and define $r$ as the distance (in the direction shown
by the arrow) from the surface. It is easy to see that $\phi$ ranges from $\tan^{-1} (a)$
to $\pi - \tan^{-1}(a)$ and $r$ ranges from $-\infty$ to $\infty$. 
The Jacobian of the transformation can be easily found to be given by
$r+f(\phi)$, where 
\beq
f(\phi) = \frac{Ra\cosec^3 \phi}{(1-a^2\cot^2\phi)^{3/2}}~.
\eeq

We can now define the mass term $m({\bf p})$ in terms of  $p_r$,$p_\phi$
and $p_x$ -  $i.e.$,  $m({\bf p}) = m_r +m_\phi+m_x$ with 
\bea  
m_r  &=&  m_0  - m_2[\partial_r^2 + (r+f)^{-1}\partial_r], \nonumber \\
m_\phi &=&  - m_2[(r+f)^{-2}(\partial_\phi^2 - (r+f)^{-}
(\partial_\phi f) \partial_\phi), \nonumber \\
{\rm and} ~~m_x 
&=& m_2p_x^2~.
\eea
The bulk Hamiltonian $H=m({\bf{p}})\tau_z + (A_xp_x\sigma_x + A_yp_y\sigma_y + A_zp_z\sigma_z)\tau_x$ can now be written 
in terms of the curvilinear coordinates $(r,\phi,x)$
as 
\beq
H_{\rm bulk} =  H_r + H_\phi +H_x  \nonumber 
\eeq
with  
\bea
 H_r &=&  \left(\begin{array}{cc} m_r\tau_z + iA_z\sin(\phi)\partial_r\tau_x & A_y\cos(\phi)\partial_r\tau_x \\ 
		-A_y\cos(\phi)\partial_r\tau_x & m_r\tau_z - iA_z\sin(\phi)\partial_r\tau_x \end{array}\right), \nonumber \\
 H_{\phi} &=& \left(\begin{array}{cc} m_{\phi}\tau_z + iB_z \tau_x & -B_y\tau_x  \\
 B_y\tau_x & m_{\phi}\tau_z - iB_z\tau_x \end{array}\right), ~~{\rm and} \nonumber \\
 H_x &=& \left(\begin{array}{cc} m_x\tau_z & A_xp_x\tau_x \\ A_xp_x\tau_x & m_x\tau_z \end{array}\right)~.
 \eea
 We have used $B_z=A_z (r+f)^{-1} \cos(\phi) \partial_{\phi}$ and $B_y=A_y (r+f)^{-1} \sin(\phi) \partial_{\phi}$
in $H_\phi$ above.
Note that for a rectangular hyperbolic surface with $a=1$, our surface and expressions are equivalent to 
to those given in Ref.\cite{takane}, albeit rotated by $\pi/4$ about the $x$-axis.

To derive the effective 2D surface $(\phi,x)$ Hamiltonian from the above bulk Hamiltonian, we need to 
first solve the radial equation $H_r |\psi\rangle = E_r |\psi\rangle$.  Following Refs.\cite{takane,imura1},
we obtain  a solution of the form
$|\psi\rangle = e^{-\kappa r}|u\rangle$, where $\kappa^{-1}$ measures penetration into the bulk, provided that
we assume that $(r+f)^{-1} = \langle(r+f)^{-1}\rangle \equiv \Gamma^{-1} $ in $H_r$ where the average value $\Gamma^{-1}$ 
is defined later. 
As shown in Ref.\cite{imura1}, the boundary condition  of $\psi({\bf r} = 0) = 0$ holds when we
choose $E_r=0$.
This  gives us the values of  $\kappa$   as
\bea
\kappa_{\pm} &=&   \frac{\tilde{A}_{\phi} \pm \sqrt{\tilde{A}^2_{\phi} + 4m_0m_2}}{-2m_2} 
\nonumber \\
&=& \frac{\tilde{A}_{\phi}}{-2m_2}\left[1 \pm \sqrt{1 + \frac{4m_0m_2}{\tilde{A}^2_{\phi}}}\right]
\eea
where
$ \tilde{A}_{\phi} \equiv  A_\phi  - \Gamma^{-1} m_2 $ and $A_{\phi} = \sqrt{A_y^2 \cos^2(\phi) + A_z^2 \sin^2(\phi)} $. 
We can also get solutions with $\tilde \kappa$
being the negative of the expressions on the RHS. 
But since it is only the positive values which are compatible with the boundary condition that
the states are localized near the surface, we focus our attention only on the 
positive solutions $\kappa_{\pm}$. To find the eigenvectors, we note that
\bea
&&\left(\begin{array}{cc}  \tau_z & 0 \\ 0 & \tau_z \end{array}\right) H_r(K)  = \nonumber \\ && 
		\left(\begin{array}{cc}m_r(K) + A_z\sin(\phi)K\tau_y & -iA_y\cos(\phi)K\tau_y \\
		iA_y\cos(\phi)K\tau_y & m_r(K) - A_z\sin(\phi)K\tau_y \end{array}\right) \nonumber \\
 && = m_r(K)\mathbb{I} + K(A_z\sin(\phi)\sigma_z + A_y\cos(\phi)\sigma_y)\tau_y \nonumber \\
&& = m_r(K)\mathbb{I} + A_{\phi}K\left(\sin(\tilde{\phi})\sigma_z + \cos(\tilde{\phi})\sigma_y\right)\tau_y~, 
\eea
where in the last line we have used $e^{i\tilde{\phi}} = {(A_y\cos{\phi} + iA_z\sin{\phi})}/{A_{\phi}}$.
We then obtain the two basis (normalised) eigenstates  of $H_r$ given by
$ |\psi_{\pm}\rangle = \mathcal{N_{\phi}}\rho(r,\phi)|u\pm\rangle$  where   $\rho(r, \phi) = 
e^{-K_{+}r} - e^{-K_{-}r}$,
\beq
|u\pm> = \frac{1}{\sqrt{2}} \left(\begin{array}{c}\sin\left(\frac{\tilde{\phi}}{2} \mp \frac{\pi}{4}\right) \left(\begin{array}{c} 1 \\ \pm i \end{array}\right)
			\\ i\cos\left(\frac{\tilde{\phi}}{2} \mp \frac{\pi}{4}\right) \left(\begin{array}{c} 1 \\ \pm i \end{array}\right) \end{array}\right)~,
\eeq
and $\mathcal{N_{\phi}}$ is a $\phi$ dependent normalisation factor.
The convention chosen above is such that $|\psi_+\rangle$ corresponds to the solution in which spin is pointing along the negative 
	$r$ direction (ie outside the insulator) and the spin for $|\psi_-\rangle$ points along the positive $r$ direction (inside the insulator). 
Note that the direction of the real spin is determined by the angle $\tilde\phi$, which
depends on the values of $a$, which  determines the curve of the surface and also $A_y$ 
and $A_z$ which are material dependent parameters. It also varies as a function
of $\phi$ , as we sweep the surface across the range of $\phi$. 
But it is always parallel to the surface, as is expected from spin-momentum locking.

Now, let us derive the effective 2D surface Hamiltonian in the $|\psi\pm\rangle$ space. Any surface
state $|\chi\rangle$ can be represented as $|\chi\rangle = \chi_+ |\psi_+\rangle + \chi_-|\psi_-\rangle$. We can now
define a two component spinor $ \chi = (\chi_+,\chi_-)^T$ and define the 
effective Hamiltonian for $\chi$ as
\beq
 H_{\rm eff} = \begin{pmatrix} \langle\psi_+|H_\phi+H_x|\psi_+\rangle & \langle\psi_+|H_\phi+H_x|\psi_-\rangle \\
                                                \langle\psi_-|H_\phi+H_x|\psi_+\rangle & \langle\psi_-|H_\phi+H_x|\psi_-\rangle \end{pmatrix}.
\eeq
Note that $+$ and $-$ refers here to real spin-up and spin-down,  but the quantisation axis,
and hence the meaning  of spin up and down, continuously
changes along the hyperbolic surface.  We shall 
essentially follow the same steps as in Ref.\cite{takane} to compute the effective Hamiltonian.
To simplify the notation, we define $\theta_{\pm} = {\tilde{\phi}}/{2} \mp {\pi}/{4}$.
We find that the diagonal elements $\langle\psi_\pm|H_\phi+H_x|\psi_\pm\rangle=0$ and the off-diagonal elements 
are given by 
\beq
\langle\psi_\pm|H_\phi+H_x|\psi_\mp\rangle =   \tilde{\mathfrak{H}}_\pm , 
\eeq
where 
\bea
&&\tilde{\mathfrak{H}}_{\pm} = \mp \frac{\tilde{\mathcal{A}}_{\phi}}{\langle r \rangle + f} \partial_{\phi} 
				\mp \frac{1}{2} \partial_{\phi} \left(  \frac{\tilde{\mathcal{A}}_{\phi}}{\langle r 
\rangle + f} \right) - A_xp_x  \nonumber  \\
\text{and } && \tilde{\mathcal{A}}_{\phi} = \left[A_{\phi} - m_2 \Gamma^{-1} \right] \left(\partial_{\phi} \tilde{\phi} \right)~. \nonumber
\eea
In the above equations, we have used
\bea
\int_0^{\infty} dr\left[\mathcal{N}_{\phi} \rho(r,\phi) \right]^2
			  &=& \frac{\int_0^{\infty} dr \rho^2(r, \phi)}{\int_0^{\infty} dr (r+f) \rho^2(r, \phi)} 
\nonumber \\
				&=& \frac{1}{\langle r \rangle + f} 
\eea
  where 
\beq
				\langle r \rangle = \frac{\int_0^{\infty} dr r \rho^2(r, \phi)}{\int_0^{\infty} dr \rho^2(r, \phi)}~.
\eeq
We have also used 
\beq
\Gamma^{-1} = \frac {\int dr (r+f)^{-1} \rho^2}{\int dr \rho^2} 
\eeq
and 
\bea
&&\int_0^{\infty} dr \mathcal{N}_{\phi} \rho(r,\phi) \partial_{\phi} \left(  \mathcal{N}_{\phi} \rho(r,\phi) \right) \nonumber \\
&&				=	\frac{1}{2} \partial_{\phi} \left( \int_0^{\infty} dr\left[\mathcal{N}_{\phi} \rho(r,\phi) \right]^2 \right) 
			  = \frac{1}{2} \partial_{\phi} \left( \frac{1}{\langle r \rangle + f} \right). 
\eea
		
As was done in Ref.\cite{takane}, it is convenient to define a length variable, instead of the angle
variable $\phi$ as 
\beq
l = \int_{\frac{\pi}{2}}^{\phi} d\phi' \langle r \rangle(\phi') + f(\phi')
\eeq
located just below the geometric surface. The limits of $\phi \rightarrow \tan^{-1} a, \pi-\tan^{-1}a$
correspond to $l\rightarrow -\infty,+\infty$. Since the probability density must not change during
this transformation, ${\tilde\chi}_\pm (l)$ is related to $\chi_\pm(\phi)$ as
\beq
{\tilde\chi}(l) = \frac{{\chi}(\phi(l))}{\sqrt{\langle r \rangle + f}}~.
\eeq
Hence, we can solve the eigenvalue equation $H_{\rm eff} \chi_{\pm} = E \chi_{\pm}$
on the entire hyperbolic surface, by transforming $\tilde{\mathfrak{H}}_{\pm} {\chi}(\phi) = E {\chi}(\phi)$ to
\beq
\mathfrak{H}_{\pm} {\tilde\chi}(l) = E{\tilde \chi}(l)~, 
\eeq
 where $\mathfrak{H}_{\pm}$ is now given by 
\beq
\mathfrak{H}_{\pm} = \mp \tilde{\mathcal{A}}_{l} \partial_{l} \mp \frac{1}{2} \partial_{l} \left(  \tilde{\mathcal{A}}_{l}\right) - A_xp_x~,
\label{curlyham} \eeq
and  $\tilde{\mathcal{A}}_{l} $ can be considered the effective velocity of the electron along the coordinate $l$ which 
is effectively increased due to the effect of the Jacobian dependent second term in $\tilde {\mathcal{A}}_\phi$.
We focus on eigenstates with energy $E=\sqrt{(Ak)^2+(A_xk_x)^2}$ and obtain the
two surface solutions as
\beq
{{\tilde \chi}}_{\pm} = \sqrt{\frac{1}{2\tilde{\mathcal{A}}_l}} 
			\left(\begin{array}{c} 1 \\ -e^{\mp i \zeta} \end{array}\right) e^{\pm i \int^l dl' \frac{Ak}{\tilde{\mathcal{A}}_{l'}} + ik_x x}  \label{wfncs}
\eeq
	The factor of $\sqrt{2}$ above ensures that the probablity density current in the direction of $l$ is $\pm \frac{Ak}{E}$ as it should be for a free Dirac particle 
	of momentum $k$ and energy $E$. Hence, as expected, we get surface states satisfying the Dirac
equation. However, unlike on a planar surface, the definition of $Ak$ changes as we change $l$.
Since these wave-functions are valid everywhere on the hyperbolic surface, it is clear that no
backward scattering takes place anywhere and that the transmission is unity along any path
independent of the shape and size of the curvature of the surface.  Thus, we generalise the
earlier result \cite{takane} which concluded that there was no reflection at a $90^o$ corner
and find that there is no reflection even when the corner has any angle other than $90^o$.

\section{Density of states and tunneling current for polarised STM}

\subsection{Density of states}

Although there is no reflection  on the curved surface,
the density of states does change as a function of both the curvature and the sharpness of the corners.
The local density of states (DOS) is defined as
\beq
\rho_s = \int d\omega [f(\omega) - f(\omega+eV)] \sum_\nu \delta (E_\nu - eV) |{\tilde\chi}_{\pm}|^2
\eeq
and is clearly a function of the curvature parameter $a$ and the sharpness parameter $R$
through the wave-function ${\tilde\chi}_\pm$.
We will restrict ourselves to low biases and low temperatures and normalize the DOS by its value
at  $\phi = \tan^{-1}a$.  In Figs. (2a) and (2b), we show the DOS as a function of the
angle parameter $\phi$ which spans the surface. Note that the DOS shows a dip at $\phi=\pi/2$ (or $l=0$), which is the point
of maximum curvature, both for fixed $R$ in Fig. (2a) as well as for fixed $a$ in Fig. (2b). 
Note also that in Fig. (2a), the range of $\phi$ depends on the curvature
parameter $a$ and increases as the curvature increases, whereas in Fig. (2b), the range is fixed from
$\tan^{-1} a = \pi/4$ to $\pi - \tan^{-1}a = 3\pi/4$ for $a=1$. From both the curves, it is clear that
in the limit of a planar surface ($a \rightarrow \infty$ or $R\rightarrow \infty$), the DOS is flat, as is expected
for the usual planar TIs.

\begin{figure}[h]
\includegraphics[scale=0.4]{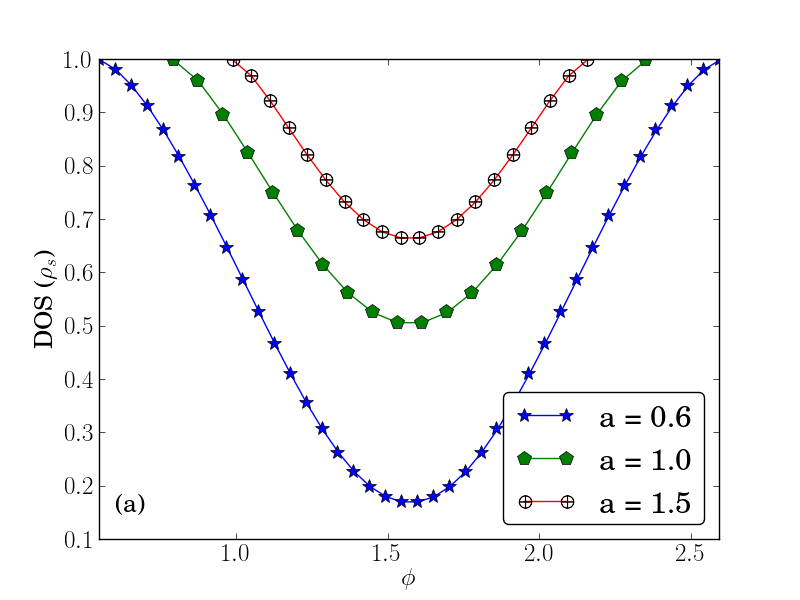}
\includegraphics[scale=0.4]{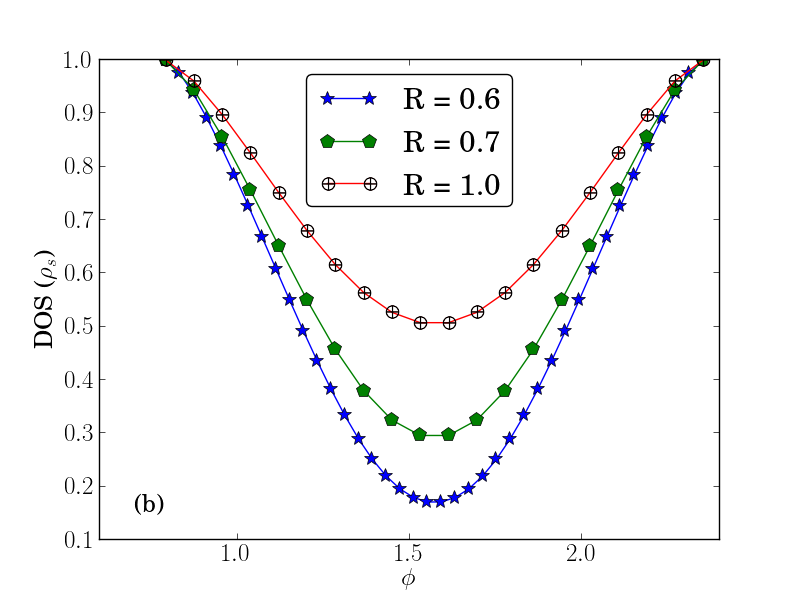}
\caption{The DOS as a function of the angle parameter $\phi$ along the surface, for fixed sharpness $R=1$ 
and for three different curvatures, normalised to unity at $\phi=\tan^{-1}a$, 
is shown in Fig. (2a), while in Fig. (2b), it is shown for three different values of the sharpness parameter and for fixed curvature $a=1$.
}\label{twoplots}
	\end{figure}

\subsection{Tunneling current for a polarized STM tip}

We now use the theory for the tunneling current of Dirac electrons on the surface of a TI, 
developed in Ref.\cite{krishnendu}.  By using the Bardeen tunneling formula\cite{bardeen},
and explicitly computing the matrix element for overlap between the tip and the material
wave-functions, they found that the tunneling current  at zero temperature was given by 
\bea
I(V) &=& I_0 |c_0|^2 \rho_t[ \rho_d +\rho_z \cos\alpha +\rho_m\sin\alpha]     \nonumber \\
    & \times & \int  d\omega [f(\omega) - f(\omega+eV)]  \nonumber \\
    {\rm with } ~~\rho_d &=&   \sum_\nu\delta(E_\nu -eV) | \psi_{\uparrow\nu}|^2 (1+\kappa_\nu^2)  \nonumber \\
     \rho_z &=&   \sum_\nu\delta(E_\nu -eV) | \psi_{\uparrow\nu}|^2 (1 -\kappa_\nu^2)  \nonumber \\
      \rho_m &=&   \sum_\nu\delta(E_\nu -eV) | \psi_{\uparrow\nu}|^2 \kappa_\nu \cos(\beta -\eta_\nu)~.
\label{conductance}
\eea
Here $c_0$ is a constant which depends on details of the tip wave-function and $\rho_t$ is
the density of states for the tip electron, assumed to be constant.
However this was derived using a flat surface for the topological insulator with the spin quantisation
axis of the electrons fixed to be in the ${\hat z}$-direction. In this basis, the tip wave-function was
represented as 
\beq
\psi_t = \begin{pmatrix}{}\cos(\alpha/2)\\~~~~\sin(\alpha/2)e^{i\beta}\end{pmatrix}
\label{tipspinor}
\eeq
and the electron wave-function on the surface was given by
\beq
\psi_\nu({\bf r};z) = \begin{pmatrix} \psi_{\uparrow\nu}({\bf r};z) \\ \psi_{\downarrow\nu}({\bf r};z) \end{pmatrix}
=\begin{pmatrix} 1\\ \kappa_\nu e^{-i\eta_\nu} \end{pmatrix}\psi_{\uparrow\nu}~. \label{tiwfn}
 \eeq
But for the hyperbolic surface, the spin quantisation axis is continuously changing as a function
of the curvature - i.e., as a function of the angle $\phi$, due to spin-momentum locking.  Hence, to express the tip wave-function
in the basis of the electron spin quantisation axis, we need to rotate the spinor by $\phi-\pi/2$  and obtain
it in the $\phi$-basis, the basis which is perpendicular to the surface.  In this basis, the tip
wave-function is given by
\beq
\psi_t^\phi = \begin{pmatrix}{} C \cos(\alpha/2) - iS\sin(\alpha/2)e^{i\beta}\\  
-iS \cos(\alpha/2) + C\sin(\alpha/2)e^{i\beta}
 \end{pmatrix} \label{tipspinorphi}
\eeq
where $C=\cos(\phi/2-\pi/4) $ and $S=\sin(\phi/2-\pi/4)$. Note that
we have rotated the basis in the counter-clockwise direction about the $x$-axis, and have
chosen the $\phi$-axis to be pointing out of the TI - $i.e.$, opposite
the normal direction defined by $r$.

Now to be able to use
the expression for the current given in Eq.\ref{conductance}, we need to 
rewrite the spinor given in Eq.\ref{tipspinorphi} in the form given in
Eq.\ref{tipspinor}.  We find that $\psi_t^\phi$ can be written as
\beq
\psi_t^\phi = e^{i\xi}\begin{pmatrix}{}\cos({\tilde\alpha/}2)\\  ~~~~\sin({\tilde\alpha}/2)
e^{i{\tilde\beta}} \end{pmatrix}
\eeq
where 
\bea
\cos{\tilde\alpha} &=& -\cos{\phi}\sin{\alpha}\sin{\beta} + \sin\phi\cos{\alpha}, \nonumber \\
\tan{\tilde\beta} &=& \tan\beta\sin\phi +\cos\phi\sec\beta\cot\alpha
\eea
and the overall (unimportant) phase of the spinor is given by
\beq
\tan\xi = -\frac{\tan(\frac{\alpha}{2}) \cos(\beta) \tan(\frac{\phi}{2} - \frac{\pi}{4})}{1 + \tan(\frac{\alpha}{2}) \sin(\beta) \tan(\frac{\phi}{2} - \frac{\pi}{4})}~.
\eeq

The wave-function of the electron on the curved surface has been derived in Eq.\ref{wfncs}. 
Comparing with the wave-function in Eq.\ref{tiwfn}, we see that $\kappa_\nu =-1$ and $\eta_\nu = \pm\zeta$.
Now, we can use the expressions given in Eq.\ref{conductance} to compute
the tunneling  conductance as 
\bea
G &=& G_0 |c_0|^2\rho_t[ \rho_d +\rho_z \cos{\tilde \alpha} +\rho_m\sin{\tilde \alpha}] \nonumber \\
{\rm with} ~~    \rho_d &=&  {\frac{1}{2{\tilde A}}} \sum_\nu\delta(E_\nu -eV),   \nonumber \\
   \rho_z &=& 0 \nonumber \\
{\rm and}~~    \rho_m  &=&  - {\frac{1}{2{\tilde A}}}  \sum_\nu\delta(E_\nu -eV) \cos({\tilde\beta} \mp \zeta)~. \label{tcond}
\eea
where $G_0=2e^2/h$.  $\rho_z=0$ because  $\kappa_\nu =-1$ , which is a result of spin-momentum locking,
due to which  the spin of the electron lies along the surface perpendicular to the quantization axis.
However, $\rho_d$ and $\rho_m $ are non-zero and both show non-trivial dependence on $\phi$ (or $l$).
In Fig.(3), the tunneling conductance   has been plotted as a function of $l$ for various values of the curvature $a$ with fixed 
sharpness parameter $R$ (for fixed tip parameters $\alpha$ and $\beta$).  Note that unlike the case for a flat surface for which the STM conductance  is constant,
(reproduced here in the large $a$ limit),  here
the STM conductance  varies as it spans the curved surface and shows a dip precisely at $l=0$. This dip depends
on the curvature and increases as the curvature increases.
\begin{figure}[h]
\includegraphics[scale=0.4]{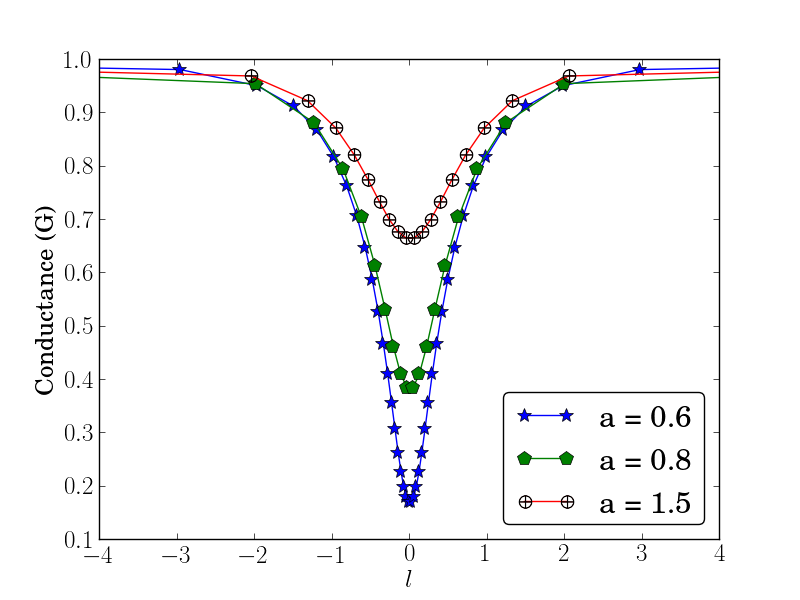}
\caption{Tunneling conductance $G$ (normalised to unity at $l\rightarrow -\infty$) 
for fixed polar and azimuthal angles 
($\alpha = 0$ and
$\beta = 0$),
as a function of $l$, the parameter along the surface, for $R=1$ and  for three   different curvatures.}\label{phi plot}
	\end{figure}

 Although $\rho_d$ in Eq.\ref{tcond} is independent of $\tilde \alpha$
and $\tilde\beta$, $\rho_m$ is dependent on both and hence shows  a non-trivial
dependence on the polar and azimuthal angles of the STM tip. Hence, the tunneling
conductance also has a non-trivial dependence on both $\alpha$ and $\beta$, which
has been shown in Figs. (4a) and (4b) respectively. As can be seen from  the plots, the conductance
shows a dip, not only at $\phi=\pi/2$ ( or $l=0$ for $a\ne 1$), but also at $\alpha=0,\pi$ in Fig.(4a)
and at $\beta=\pi/2,3\pi/2$ in Fig. (4b).
\begin{figure}[h]
\includegraphics[scale=0.4]{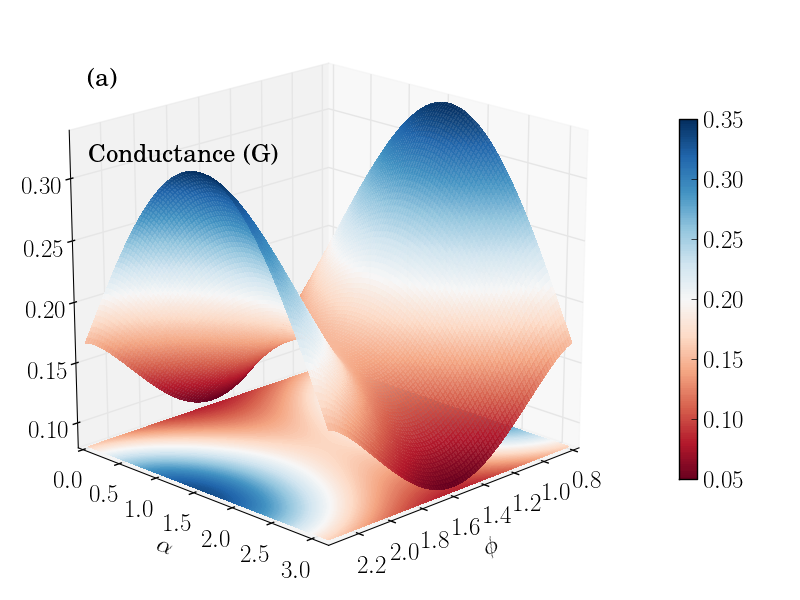}
\includegraphics[scale=0.4]{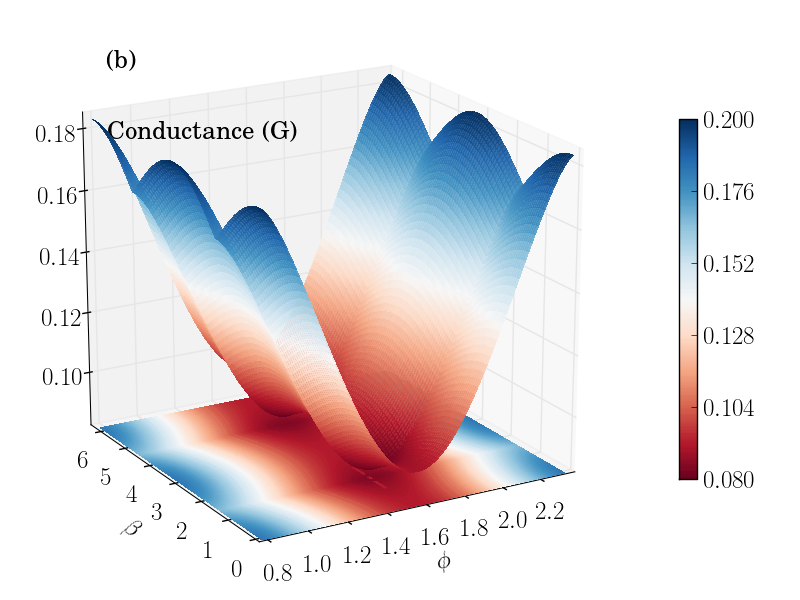}
\caption{In Fig. (4a), the tunneling conductance $G$ (normalised to unity at $\phi=\tan^{-1}a$)
is plotted as a function of
the polar angle $\alpha$  of the tip and $\phi$, the angle that spans the surface,  for  fixed  azimuthal angle
$\beta = 0 $, fixed curvature $a=1$ and fixed sharpness parameter $R=1$.
In Fig. (4b), the tunneling conductance $G$ is plotted as a function of
the azimuthal angle  $\beta$  and $\phi$,  for fixed polar angle
$\alpha = 0.1$ for the same  curvature and sharpness.
}\label{twoplotsstm}
	\end{figure}

\section{Discussion and conclusion}

Although we have restricted ourselves to concave hyperbolae for the above calculations, they
can be extended to cases where the convex side is filled with the TI. As explained in Ref.\cite{takane},
in this case, $r<0$. But the Jacobian becomes ill-defined when $|r| \gtrsim R$ due to the presence 
of the term $(-|r|+f)^{-1}$. Hence, we need to restrict ourselves to  $|r| \lesssim R$. In that case, 
a very similar analysis works and the
surface is described by the effective Hamiltonian in Eq.\ref{curlyham} with the difference
that now $\tilde{\mathcal{A}}_{\phi} = \left[A_{\phi} - m_2 \langle (-|r| + f)^{-1} \rangle \right] \left(\partial_{\phi} \tilde{\phi} \right)$. The sign change compared with the earlier case implies that the velocity of the electrons 
is now reduced, and hence its amplitude and consequently, the DOS shows peaking at the corners instead of dips. This can
also be measured by an STM tip.

In conclusion, we have found the effective Hamiltonian on curved hyperbolic surfaces and shown
that the Hamiltonian can be written in terms of a continuous coordinate which varies along
the surface. This clearly indicates that there is no back-scattering for any curvature. The sharpness
of the edge can also be changed continuously with no backscattering even
at sharp edges. However, the DOS does change as we span the surface and there is 
a dip (or peaking) of the DOS at the edge for concave (convex) surfaces. We have
also shown that the STM spectra of the Dirac electrons on  the curved surface, as measured
by a magnetized tip, shows unconventional and non-trivial dependence, not only on the
parameter spanning the surface, but also on the polar and azimuthal angles of the tip.
These measurements would provide clear evidence for the curvature of the surface.

\section*{Acknowledgments}
One of us (S.R.)  would like to thank Sourin Das, Venkat Pai, Arijit Saha, Diptiman Sen, Vijay Shenoy and Abhiram Soori for useful
discussions. We would particularly like to thank Arijit Saha for bringing Ref.\cite{takane} to our attention.

\end{document}